# Femtosecond study of $A_{1g}$ phonons in the strong 3D topological insulators: From pump-probe to coherent control

Jianbo Hu[1,2,*], Kyushiro Igarashi[3], Takao Sasagawa[3], Kazutaka G. Nakamura[3], and Oleg V. Misochko[4,5,*]

[1]State Key Laboratory for Environment-Friendly Energy Materials, Southwest University of Science and Technology, Mianyang, Sichuan 621010, China

[2]Institute of Fluid Physics, China Academy of Engineering Physics, Mianyang, Sichuan 621900, China

[3]Materials and Structures Laboratory, Tokyo Institute of Technology, R3-10, 4259 Nagatsuta, Yokohama 226-8503, Japan

[4]Moscow Institute of Physics and Technology (State University), 141700 Dolgoprudny, Moscow Region, Russia

[5]Institute of Solid State Physics, Russian Academy of Sciences, 142432 Chernogolovka, Moscow region, Russia

*To whom correspondence should be addressed. Email: jianbo.hu@caep.cn (J.H.), misochko@issp.ac.ru (O.V.M.)

Fully symmetric $A_{1g}$ phonons are expected to play a dominant role in electron scattering in strong topological insulators (TIs), thus limiting the ballistic transport of future electronic devices. Here, we report on femtosecond time-resolved observation of a pair of $A_{1g}$ coherent phonons and their optical control in two strong 3D TIs, $Bi_2Te_3$ and $Bi_2Se_3$, by using a second pump pulse in ultrafast spectroscopy measurements. Along with well-defined phonon properties such as frequency and lifetime, an obvious phonon chirp has been observed, implying a strong coupling between photo-carriers and lattices. The coherent phonon manipulation, on the other hand, allows us to change the phonon amplitude selectively but does not affect either the frequency or coherence lifetime of the chosen mode.

Strong topological insulators (TIs) are a new class of quantum matter for which the bulk phases are ordinary band insulators but possess nontrivial conducting surface with nondegenerate spins.[1,2] These surface states (SS) shaped in an odd number of Dirac cones are topologically protected against back scattering and are immune to defects as long as the disorder does not violate time reversal symmetry. A particularly unusual SS property is that its electrons' spin-orientations are locked perpendicular to their momenta that reduces the phase space for spin-conserving scattering.[1,2] However, there still remain processes that limit ballistic transport of the SS electrons. Among them, those driven by electron-phonon interactions in particular have been the subject of intense study because they affect any finite-temperature application. Indeed, future technical improvements may eventually eliminate surface defects, but phonons will be always present. Consequently, coupling to phonons should be the dominant SS scattering mechanism at finite temperatures. A number of measurements seem to have arrived at a consensus that electron scattering in strong TIs is dominated by fully symmetric $A_{1g}$ phonons.[3-6] Ultrafast time-resolved reflectivity experiments have observed coherent oscillations assigned to the $A_{1g}$ phonons[7-16] but because such experiments cannot distinguish electronic bands, it is still not clear whether these oscillations involve the SS. Despite many excellent works concerning coherent lattice dynamics,[3-16] phonons and electron-phonon interactions, both critical toward applications based upon TIs, are still not well understood. In this regard, coherent control is a powerful and sensitive technique[17] to study phonon peculiarities. In this Letter we report a time-domain experiment on $Bi_2Te_3$ and $Bi_2Se_3$, which reveals that both $A_{1g}$ optical phonons in each of the two typical TIs with a single Dirac cone can be coherently controlled. The coherent phonon manipulation for the fully symmetric modes allows to change the phonon amplitude selectively but does not affect either the frequency or coherence lifetime of the chosen mode.

Bi$_2$Te$_3$ and Bi$_2$Se$_3$ are 3D layered materials, both known historically as excellent thermoelectric materials with diverse applications. They have been shown to be ideal candidates for studying room temperature topological insulating behavior as having a topologically nontrivial band gap of up to 0.3 eV, which is much larger than the room temperature energy scale. Therefore, these binary compounds are attractive for both fundamental research and potential applications in spintronics, quantum computing, and low-energy dissipation electronics. Both compounds share the same rhombohedral crystal structure belonging to the space group $D_{d5}^{3}$, composed of hexagonal close-packed atomic layers periodically arranged along the trigonal $c$ axis.[18] The atomic arrangement can be considered as repeating units with each consisting of five atomic –Te$^1$(Se$^1$)-Bi-Te$^2$(Se$^2$)-Bi-Te$^1$(Se$^1$) – layers called quintuple layers (QLs). The primitive unit cell contains five atoms, hence there are 15 lattice modes at the center of the Brillouin zone, two of which are fully symmetric phonons.[18]

Samples of Bi$_2$Se$_3$ and Bi$_2$Te$_3$ were synthesized using conventional Bridgman method.[12,13,15] All measurements were performed at room temperature on freshly cleaved crystals. Our ultrafast pump-probe setups are discussed in detail elsewhere.[12,16] For this experiment, we use a Ti:Sapphire laser operating at 86 MHz repetition rate to excite the samples using pulses of duration $\tau_p$=40 fs and photon energy 1.5 eV, and subsequently probe the modified reflectivity $\left(\frac{\Delta R}{R_0}\right)$ using an isotropic detection scheme with weaker, delayed pulses. The probe power is always one order of magnitude weaker than the pump power which provides typical fluence of 0.1 mJ/cm$^2$. For coherent control experiment, our pump arm is essentially a Michelson-type interferometer, providing collinear pairs of excitation pulses, whose interpulse separation Δt can be scanned by a linear motorized stage with a time resolution better than 1 fs.

A representative time-resolved transient reflectivity from $Bi_2Te_3$ for a single pulse excitation is shown in Fig.1(a). At time zero ultrafast pulse drives a direct transition from the valence band to high-lying unoccupied bulk and surface states, and these electrons scatter down to the SS and conduction band. These events are responsible for a non-oscillatory signal superimposed on which are oscillations. The oscillations are induced by the modulation of dielectric constant due to the synchronized ion motion. During pump pulse duration, few scattering processes have occurred and the lattice wave function may be described by a superposition of different lattice states with well-defined phase relations resulting in quantum beats which are a signature of phase coherence.[17] The decay of coherent phonons can be monitored directly by measuring the damping rate of oscillations. This rate is a measure of how strongly the particular lattice mode is coupled to the bath (electrons plus other phonon modes).

In Fig.1(a) obvious beatings can be observed demonstrating the coexistence of two modes with close but distinct frequencies. The beatings exist only at small time delays disappearing at longer time delays. We begin our analysis of the transient reflectivity by making fast Fourier transform (FFT) of the signal in the whole scanned range, the result of which is shown in Fig.1(b). It presents FFT spectrum with two well resolved peaks, whose frequencies with their assignments are listed in Table I. The lattice displacements of these two fully symmetric phonon modes are displayed in the inserts of Fig.2(b). The $A_{1g}^{(I)}$ vibrations occur at lower frequency than $A_{1g}^{(II)}$. The latter mode, where the outer Bi and $Te^1$ atoms move out of phase, is mainly affected by the forces between Bi and $Te^1$ atoms. In the low-frequency $A_{1g}^{(I)}$ mode, the outer Bi–$Te^1$ pairs move in phase. The peak height in the FFT spectrum is almost 3 times larger for the low-frequency $A_{1g}^{(I)}$ mode, however, the ratio for the integrated amplitudes is significantly smaller because the linewidth for the high-frequency mode appreciably exceeds that for the low-frequency one. The faster decay for $A_{1g}^{(II)}$ mode is caused by either a stronger electron-phonon coupling, or more available channels for the anharmonic decay as compared to $A_{1g}^{(I)}$ mode. It is also interesting to notice in Fig.1(b) that the line-shapes for both modes are asymmetric being steeper at low frequency side. In the time-domain, this asymmetry signals that the coherent oscillations are negatively chirped with their frequency increasing over time as we can infer from the lineshape of their Fourier transform. The physical reason for the chirp is a Fano-like interference between the decaying electronic signal and the phonon oscillation.[18] Here, the most intriguing characteristic of the coupling between electrons and phonons is that the phonon chirp is short-lived completely disappearing after one or two oscillation cycles. This behavior suggests that the coupling with fully symmetric modes is different for the case of non-thermalized and thermalized carriers. Another reason for the short lived chirp might be due to fast carrier ambipolar diffusion away from the surface which, as reported in Ref. [11], can be very fast (on the order of 500 cm$^2$/s) and very short optical penetration depth (on the order of 10 nm).

To assess the coherent amplitude ratio, we fit the oscillatory signal in real time. To this end, we first remove the non-oscillatory background either by a digital band pass filter or by fitting and subtracting it as a sum of several exponents.[16] Then, we fit the oscillatory residual to two damped harmonic oscillations

$$\left(\frac{\Delta R}{R_0}\right)_{osc} = \sum_i A_i \exp(-\frac{t}{\tau_i}) \sin(\Omega t + \phi_i) \quad (1)$$

With this fitting procedure one can obtain the amplitude, phase, coherence life time and frequency of each fully symmetric mode. The green solid line in Fig.2(a) corresponds to the fit using Eq. (1) where the experimental points are open circles. Figure 2(b), in which the traces are shifted vertically for clarity, presents the contribution of each fully symmetric mode. Both modes are cosine like[15] and their initial amplitude ratio $A_{1g}^{(I)}/A_{1g}^{(II)}$ is slightly less than one and half. We note that the determination of the initial phase is hindered by a number of factors. The main error source is related to the determination of zero delay, which fixes the value of the initial phase. In our experiments, the zero delay is taken at the position of the cross-correlation peak which is somehow arbitrary. For $Bi_2Te_3$, an error of 20 fs in the zero delay gives a 15° (30°) error in the phase for $A_{1g}^{(I)}$ ($A_{1g}^{(II)}$) mode.

The observation of both fully symmetric modes in the $Bi_2Te_3$ transient oscillations raises the question how far they are mutually coupled and influence each other or whether they are independent. Additional insight into this coupling can be gained from coherent control experiments in which a second time-delayed pump pulse is employed to control the oscillation of a coherent phonon launched by the first pump pulse. Such control of coherent phonon oscillations in the time domain has been demonstrated in a number of materials, including selective excitation of certain phonon modes.[17,19,20] Two-pulse excitation impulsively excites a particular phonon mode twice, the second time with a chosen phase relative to the first. In this case, the first pump pulse initiates a coherent motion with all atoms moving in synchrony, whereas the second excites a phase-shifted replica of the first, and the two interfere either constructively or destructively. Thus, the final state is a lattice system with one selected mode excited to larger amplitude or almost canceled out. We are first interested in high-frequency oscillations of $Bi_2Te_3$, whose coherence time is much less than the dephasing time for low-frequency oscillations. As a result, high-frequency $A_{1g}^{(II)}$ contribution to the transient reflectivity is significant only at small delays where it is masked by strong $A_{1g}^{(I)}$ contribution. Choosing the interpulse separation $\Delta t=272$ fs, which coincides with the half period of low-frequency $A_{1g}^{(I)}$ oscillations, and equal intensity for both pump pulses we are able to completely suppress the stronger $A_{1g}^{(I)}$ contribution. This can be easily seen as the disappearance of beating pattern in the time-domain, see Fig.1(c), or, alternatively, in FFT spectrum dominated by the $A_{1g}^{(II)}$ mode for two-pulse excitation, see Fig.1(d). However, coherence time and frequency of the $A_{1g}^{(II)}$ phonons obtained after the discrimination of low-frequency $A_{1g}^{(I)}$ oscillations practically coincide with those obtained from the single-pulse experiment. Similarly, discriminating the high-frequency $A_{1g}^{(II)}$ mode with the interpulse separation of $\Delta t=125$ fs, as shown in Fig.1(e) and 1(f), we observed that $A_{1g}^{(I)}$ lifetime and frequency remain unchanged. Even though we did not observe any change in coherence time, using coherent control one can significantly improve the accuracy of determination of the lattice dynamics parameters, and create the lattice state with prescribed properties since we can selectively switch on and off a particular phonon mode.

Now we turn to coherent lattice dynamics in $Bi_2Se_3$. Similar to the $Bi_2Te_3$, it consists of a non-oscillatory background and two oscillatory components appearing right after laser excitation.[12] FFT shown in the inset of Fig. 3 reveals slow and fast oscillations at 2.14 and 5.13 THz, corresponding to the frequencies of $A_{1g}^{(I)}$ and $A_{1g}^{(II)}$ phonon modes. It can be noted that the frequencies for fully symmetric modes in $Bi_2Se_3$ are a bit larger than in $Bi_2Te_3$ because Te is heavier than Se. The parameters obtained from fitting the oscillatory part to Eq.(1) are listed in the Table 1. Similar to the observations in $Bi_2Te_3$, the damping rate is larger for the high-frequency $A_{1g}^{(II)}$ mode.

Results from three different excitation schemes for $Bi_2Se_3$ are shown in Fig.3. Setting the zero delay at the second pump arrival, Fig.3 shows the oscillatory part of the transient reflectivity for $Bi_2Se_3$ excited by two consecutive pump pulses with $\Delta t = 450$ fs. As a comparison, we also present the signals excited by each pump pulse alone. It is clear that the signal excited by the first pump pulse is modified by the second one. However, the simple sum of these two single-pump signals is just the same as the two-pump signal. Therefore, it is straightforward to understand coherent control as the superposition of two phonon wave packets excited independently by each pump pulse. The phase difference, which depends on the interpulse separation $\Delta t$, determines the interference to be either constructive or destructive. Based on the comparison, we can infer that the coherent control in our case does not affect either frequency or lifetime of coherent modes, at least in our linear excitation regime where the frequency and lifetime are independent of pump intensity.[18] Otherwise the two-pump signal should have deviated from the sum of the two single-pump signals.

By scanning the interpulse separation in the range of 450-700 fs we observe that each coherent $A_{1g}$ amplitude is modulated with the mode frequency, whereas its frequency and coherence lifetime remain unaffected, as shown in Fig. 4(a). While both modes are present at $\Delta t$ =580 fs, the amplitude of the high-frequency mode is almost completely suppressed and only low-frequency mode remains at $\Delta t$ =460 fs. The data are described nicely by a model with two independently controlled modes. Indeed, the amplitude of $A_{1g}^{(II)}$ mode is harmonically modulated by the second pump pulse with the period of ~200 fs, which is the period of coherent $A_{1g}^{(II)}$ phonons. Although our scanning range is not large enough to observe a full cycle modulation for coherent $A_{1g}^{(I)}$ phonons, we still can estimate that the amplitude modulation of $A_{1g}^{(I)}$ mode occurs at its intrinsic period close to 470 fs. For the whole scanned range, the lifetimes of two $A_{1g}^{(I)}$ and $A_{1g}^{(II)}$ coherent modes, however, exhibit nearly no change, see Fig.4(b). Also we did not observe any dependence of non-oscillatory background on the interpulse separation in either $Bi_2Te_3$ or $Bi_2Se_3$. This is at odds with the results obtained for $Bi_2Se_3$ where the damping rate for the electronic background was observed to be dependent on the coherent amplitude suggesting more efficient energy transfer when the lattice is in a coherent state.[10] Another example of coherent control of energy transfer to the lattice can be found in Ref. [21]. This discrepancy may be caused by either different excitation regime: linear vs. nonlinear, or it can arise from a different range of scanned interpulse separation. When in our case the range does not exceed 700 fs, it was significantly larger in Ref. [10]. Taking into account that electronic coherence can be present in our scanned range[22] but unlikely extend to the range explored in Ref. [10], it might be possible to understand the above mentioned disagreement.

Now let us briefly address the issue as to whether the coherent oscillations seen in our time domain experiments are caused either by SS or bulk electrons. The penetration depth for 800 nm wavelength is around 10 nm for both $Bi_2Te_3$ and $Bi_2Se_3$ samples that means that at least ten QLs are excited and probed in our experiments as the width of QL is about 1 nm. SS are localized to the surface region, and their spread into the bulk, is about 2 or 3 QLs.[23] Thus the main contribution to the coherent signal most likely comes from the bulk bands. However, a consensus about the electron-phonon coupling in both $Bi_2Te_3$ and $Bi_2Se_3$ has yet to be reached. We did not observe the phonon softening predicted by DFT calculations[6] showing the softening of $A_{1g}$ lattice distortions at the surface as compared to the bulk. The appearance of short-lived chirp in frequency for both $A_{1g}$ modes possibly demonstrates that both the bulk and surface electrons couple to these phonons. However, the question remains as to why the chirp is negative and not positive as expected from a softening at the crystal surface.

Based on the lack of phonon softening and the comparison of the optical penetration depth with the SS wave function we can suggest that the $A_{1g}$ oscillations measured in the ultrafast pump-probe experiments arise predominantly from the bulk electrons rather than from the SS. This is, of course, a rough estimate and it does not mean that the bulk and surface states contribute proportionally to the modulation of the optical reflectivity. We have demonstrated that these $A_{1g}$ optical phonons can be coherently controlled and this coherent phonon manipulation allows to change the phonon amplitude selectively but does not affect either its frequency or coherence lifetime because the second pump pulse does not modify the overlap between the coherent phonon mode and the bath. Nevertheless, by using such a mode selective excitation technique, we can easily observe a particular, even weak, phonon mode by suppressing other stronger modes. Furthermore, we can use this coherent control technique to realize the data storage and processing by setting two coherent modes as a pair of qubits distinguished by their frequency.

The authors thank Dr. Katsura Norimatsu for experimental assistance. This work was in part supported by the Russian Foundation for Basic Research (No. 17-02-00002) and by the Government of the Russian Federation (Agreement No. 05.Y09.21.0018). J.H. acknowledges the support from China 1000-Young Talents Plan.

[1] M. Z. Hasan and C. L. Kane, Rev. Mod. Phys. **82**, 3045 (2010).

[2] J. E. Moore, Nature **464**, 194 (2010).

[3] X. X. Zhu, L. Santos, C. Howard, R. Sankar, F. C. Chou, C. Chamon, and M. El-Batanouny, Phys. Rev. Lett. **108**, 185501 (2012).

[4] C. Howard and M. El-Batanouny, Phys. Rev. B **89**, 075425 (2014).

[5] M. V Costache, I. Neumann, J. F Sierra, V. Marinova, M. M Gospodinov, S. Roche, and S. O Valenzuela, Phys. Rev. Lett. **112**, 086601 (2014).

[6] J. A Sobota, S. L. Yang, D. Leuenberger, A. F Kemper, J. G Analytis, I. R Fisher, P. S Kirchmann, T. P Devereaux, and Z. X. Shen, Phys. Rev. Lett. **113**, 157401 (2014).

[7] A. Q. Wu, X. Xu, and R. Venkatasubramanian, Appl. Phys. Lett. **92**, 011108 (2008).

[8] Y. Wang, X. Xu, and R. Venkatasubramanian, Appl. Phys. Lett. **93**, 113114 (2008).

[9] N. Kamaraju, S. Kumar, and A. K. Sood, EPL **92**, 47007 (2010).

[10] J. Qi, X. Chen, W. Yu, P. Cadden-Zimansky, D. Smirnov, N. H. Tolk, I. Miotkowski, H. Cao, Y. P. Chen, Y. Wu, S. Qiao, and Z. Jiang, Appl. Phys. Lett. **97**, 182102 (2010).

[11] N. Kumar, B. A. Ruzicka, N. P. Butch, P. Syers, K. Kirshenbaum, J. Paglione, and H. Zhao, Phys. Rev. B **83**, 235306 (2011).

[12] K. G. Nakamura, J. Hu, K. Norimatsu, A. Goto, K. Igarashi, and T. Sasagawa, Solid State Commun. **152**, 902 (2012).

[13] K. Norimatsu, J. Hu, A. Goto, K. Igarashi, T. Sasagawa, and K. G. Nakamura, Solid State Commun. **157**, 58 (2013).

[14] J. Flock, T. Dekorsy, and O. V. Misochko, Appl. Phys. Lett. **105**, 011902 (2014).

[15] O. V. Misochko, A. A. Mel'nikov, S. V. Chekalin, and A. Yu. Bykov, JETP Lett. **102**, 235 (2015).

[16] O. V. Misochko, J. Flock, and T. Dekorsy, Phys. Rev. B **91**, 174303 (2015).

[17] K. Ishioka and O. V. Misochko, in *Progress in ultrafast intense laser science v*, edited by K. Yamanouchi, A. Giulietti, and K. Ledingham (Springer-Verlag Berlin Heidelberg, Berlin, 2010), pp. 23.

[18] O. V. Misochko and M. V. Lebedev, JETP **120**, 651 (2015).

[19] J. Hu, O. V. Misochko, A. Goto, and K. G. Nakamura, Phys. Rev. B **86**, 235145 (2012).

[20] J. Hu, O. V. Misochko, and K. G. Nakamura, Phys. Rev. B **84**, 224304 (2011).

[21] C. M. Liebig, Y. Wang, and X. Xu, Opt. Express **18**, 20498 (2010).

[22] D. L. Greenaway and G. Harbeke, J. Phys. Chem. Solids **26**, 1585 (1965).

[23] W. Zhang, R. Yu, H. Zhang, X. Dai, and Z. Fang, New J. Phys. **12**, 065013 (2010).

**TABLE I**. Comparison of phonon frequencies, lifetimes and amplitudes obtained for two fully symmetric modes in binary compounds from femtosecond time-resolved spectroscopy.

| Mode / Crystal | Frequency (THz) | Lifetime (ps) | $A_{1g}^{(I)}/A_{1g}^{(II)}$ ratio |
|---|---|---|---|
| $A_{1g}^{(I)}$ $Bi_2Te_3$ | 1.84 | 3.61 | 1.42 |
| $A_{1g}^{(II)}$ $Bi_2Te_3$ | 4.01 | 1.19 | |
| $A_{1g}^{(I)}$ $Bi_2Se_3$ | 2.09 | 3.22 | 0.91 |
| $A_{1g}^{(II)}$ $Bi_2Se_3$ | 5.20 | 1.09 | |

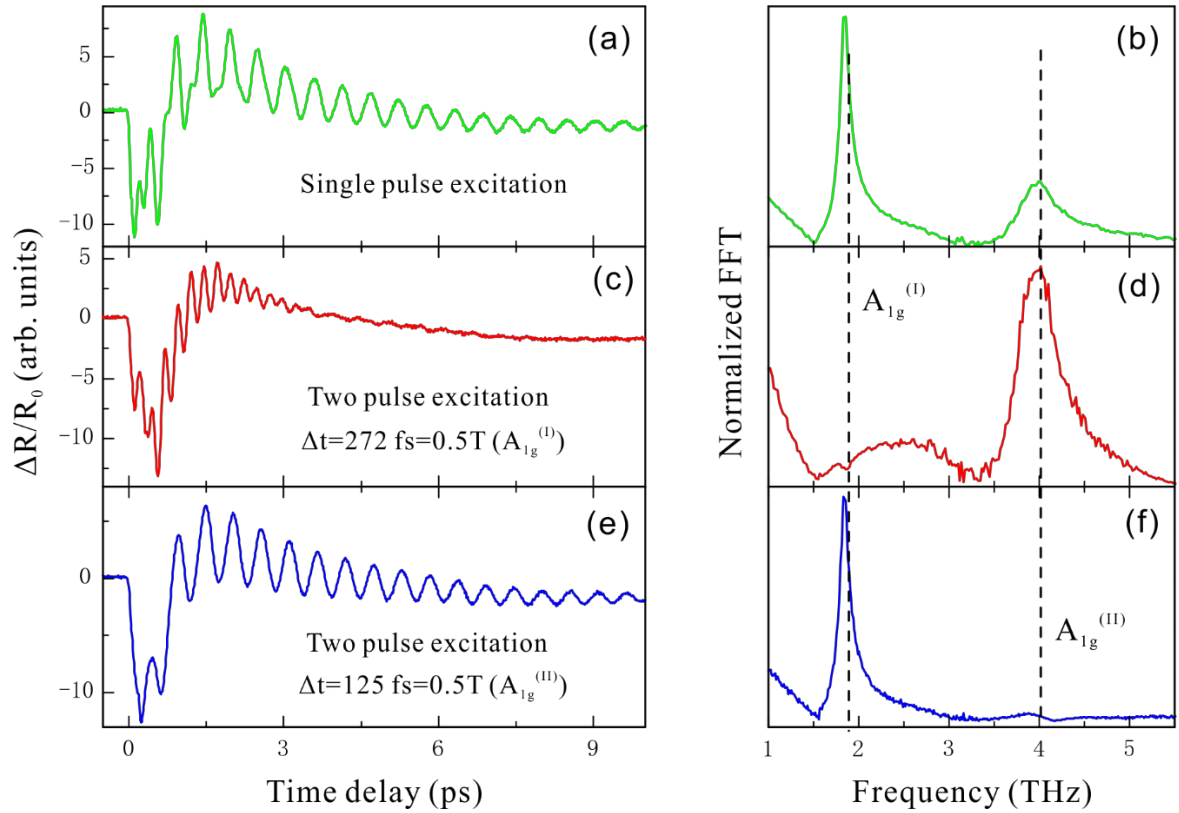

FIG.1 Transient reflectivity of $Bi_2Te_3$ and its FFT for different excitations schemes described in the text.

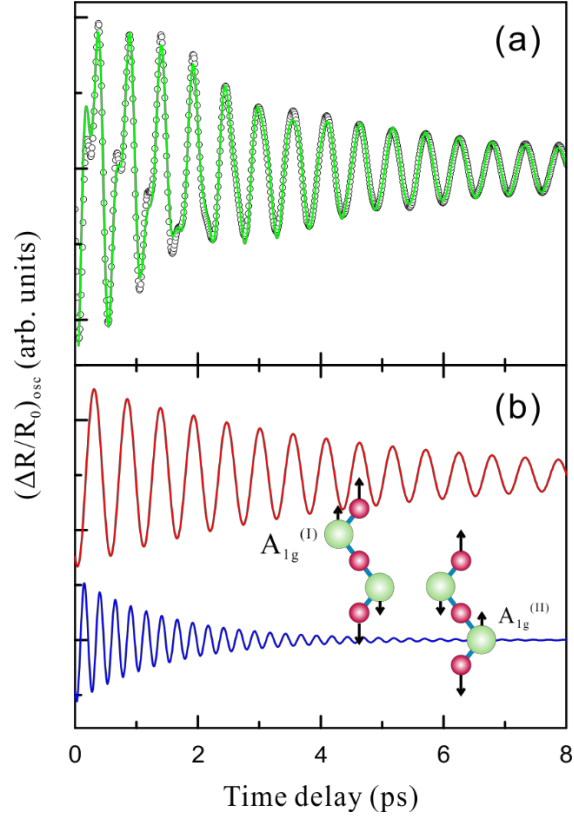

FIG.2. (a) Oscillatory part of transient reflectivity for $Bi_2Te_3$ for single pulse excitation and its decomposition to two frequency components (b). The insets in the bottom panel (b) display the eigenvectors of two $A_{1g}$ phonon modes. The green solid line corresponds to the fit using Eq. (1) where the experimental results are open circles.

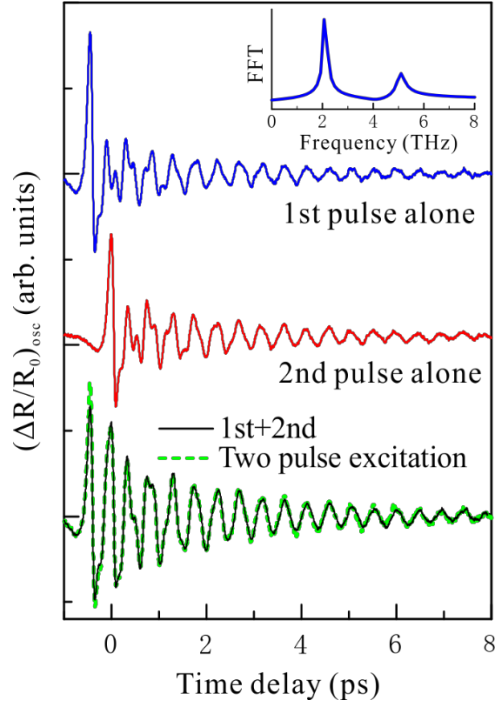

FIG. 3. Oscillatory part of the time-domain signals for $Bi_2Se_3$ excited by either single pump pulse (blue and red solid lines) or two consecutive pump pulses (green dashed line). The sum of two single-pump signals is shown by the black solid line. The inset shows the FFT for the first pulse excitation. Here, the interpulse separation is 450 fs and the traces are offset vertically for clarity.

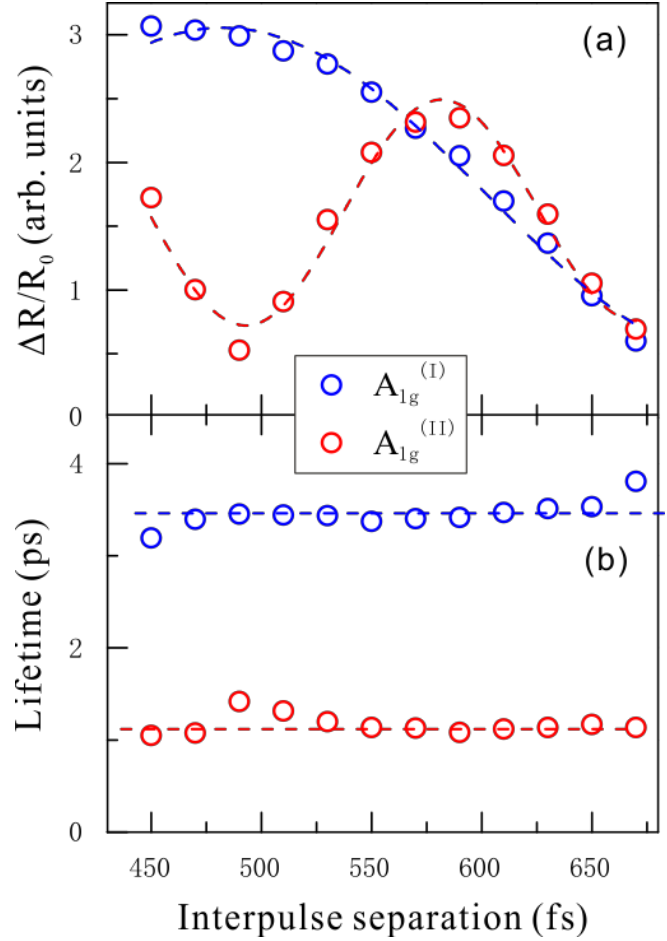

FIG.4. Coherent amplitude (a) and lifetime (b) as a function of interpulse separation. The dashed lines in (a) are fits to $\frac{A}{2}[1+\cos(\frac{2\pi}{T}\Delta t)]$ where T is the phonon period and A is the amplitude for excitation with both coinciding in time pump pulses (T=467 fs for $A_{1g}^{(I)}$ and T=196 fs for $A_{1g}^{(II)}$). The dashed lines in (b) are just a guide to the eye.